# Deep Artifact-Free Residual Network for Single Image Super-Resolution

Hamdollah Nasrollahi[1], Kamran Farajzadeh[1], Vahid Hosseini[1], Esmaeil Zarezadeh[2], Milad Abdollahzadeh[3,4]
[1]Islamic Azad University, Science and Research Branch, Tehran, Iran
[2]Amirkabir University of Technology, Tehran, Iran
[3]Faculty of Electrical and Computer Engineering, University of Tabriz, Tabriz, Iran
[4]Singapore University of Technology and Design (SUTD), Singapore
Corresponding Author's Email: Milad.abdollahzadeh@tabrizu.ac.ir

**Abstract**

*Recently, convolutional neural networks have shown promising performance for single-image super-resolution. In this paper, we propose Deep Artifact-Free Residual (DAFR) network which uses the merits of both residual learning and usage of ground-truth image as target. Our framework uses a deep model to extract the high frequency information which are necessary for high quality image reconstruction. We use a skip-connection to feed the low-resolution image to network before the image reconstruction. In this way, we are able to use the ground-truth images as target and avoid misleading the network due to artifacts in difference image. In order to extract clean high frequency information, we train network in two steps. First step is a traditional residual learning which uses the difference image as target. Then, the trained parameters of this step are transferred to the main training in second step. Our experimental results show that the proposed method achieves better quantitative and qualitative image quality compared to the existing methods.*

## 1. Introduction

Single-image super-resolution (SR) is the procedure of reconstructing a high-resolution (HR) image from a low-resolution (LR) one. In the literature, a plethora of algorithms are proposed for SR. Early algorithms are based on the interpolation such as Bicubic interpolation, Lanczos resampling [1] and the improved ones that use the statistical image priors [2, 3]. Recently, learning-based methods have shown the state-of-the-art performance using a mapping from LR to HR image patches. The learning is performed with different algorithms including local linear regression [4, 5], dictionary learning [6] and random forest [7]. Recently, the Super-Resolution Convolutional Neural Network (SRCNN) [8] is proposed to learn an end-to-end, non-linear mapping from LR to HR images. Several extensions are proposed to improve the performance of SRCNN and accelerate its training [9-11].

The more contextual information used in the network, the more we will be able to improve the performance of the network for SR task. The most efficient way to utilize more contextual information is to increase the receptive filed via using deeper structure. However, increasing the depth results in the notorious problem of exploding/vanishing gradients which hampers the network convergence [12]. One effective solution is deep residual learning framework [13]. One way to apply residual learning for SR task, is to make the network to estimate the difference between the HR image and upscaled version of LR image as high frequency details. After estimating the residual, the upscaled LR image is added to the network output to reconstruct the final HR image. This approach is effective, however, the difference between the HR image and the upscaled version of the LR image contains some artifacts. Using these artifacts as estimation target, misleads the network and degrades the quality of the reconstructed image.

In this paper, in order to simultaneously use the benefits of the residual learning and ground truth target image, we use the skip connection. In this way, we are able to feed the exact copy of the LR image to network just before image reconstruction step. The LR image is concatenated to the extracted feature maps to provide sufficient low and high frequency information for reconstruction part to estimate the HR image. Therefore, network can be trained by ground truth target without forcing it to carry the LR image from input to the output. The LR image skips a large number of intermediate layers by feeding before reconstruction and may result in "dead features" problem in these layers. In this work, we address this problem by separating the training into two steps. In the first step, we use traditional residual learning to lead the intermediate layers to reconstruct the image details. Then we transfer the trained weights and biases to the main structure and fine-tune these parameters in the second step of the training. In order to provide pleasant and sharp images, we use a more robust loss function than the $\ell_2$ loss function. The $\ell_2$ loss function fails to capture the whole multi-modal distributions of HR images. Therefore, the reconstructed images are often overly-smooth which is not pleasant for human visual system. The applied robust loss function helps our framework to provide sharper images compared to the existing methods.

The reminder of this paper is organized as follows. Section 2 discusses the related works in the literature. Section 3 ex-plains the proposed method for single image SR. The experimental results are provided in section 4. Finally, Section 5 concludes the paper.

## 2. Related Work

Single-image SR algorithms, can be categorized into four types based on the image priors: prediction models, edge-based methods, image statistical and example-based methods. The algorithms within prediction models category apply a mathematical formula on input LR image to generate the corresponding HR image [1]. The generated HR images by these algorithms have good smooth regions, however in the high frequency regions the gradients are insufficient. In edge-based methods [14], image priors are learned from edge features such as the width and depth of an edge or the parameter of gradient profile. Then, these priors are used to produce the HR image. The produced HR images with these algorithms have sharp and clean edges. However, they are less effective for the reconstruction of complex structures like textures. In image statistical methods [2, 3], image priors are exploited using statistical metrics like heavy-tailed gradient distribution and sparsity property of large gradients. These algorithms require a high computational complexity.

Among these four, the example-based algorithms have shown the state-of-the-art results [15]. In internal example-based algorithms, the self-similarity in images are utilized to construct LR-HR patch pairs [16-18]. More relevant training patches are available in internal example-based algorithms; however, the number of LR-HR patch pairs are not sufficient for textural variation to be covered. In external example-based methods a LR-HR patch mapping is learned using external datasets. The way of learning the compact dictionary or manifold space to relate the LR-HR patches is the main difference between these methods. Several supervised learning methods are used such as manifold embedding [19], nearest neighbor [20], kernel ridge regression [3] and sparse representation [21]. Nowadays, the sparse-coding-based SR algorithm and its extensions [4, 22] are among the state-of-the-art. In these algorithms, the patches are the focus of optimization. Patch extraction and aggregation are considered as pre/post-processing steps and handle separately.

Recently, the deep convolutional neural networks have shown promising performance on various computer vision fields like image classification [23, 24], activity recognition [25], defect detection [26] and image retrieval [27, 28]. The super resolution convolutional neural network (SRCNN) [8] is the first deep learning based end-to-end image SR algorithm. Inspired by the sparse-coding-based algorithms, the SRCNN learns a LR-HR mapping. However, instead of modeling the LR-HR mapping in patch space, SRCNN learns the nonlinear mapping by optimizing all the steps. Due to the high similarity between the input and output, in addition to the reconstruction of the high frequency information (details), the SRCNN carries the input to the end layer. This problem limits the SRCNN to use only three layers and therefore SRCNN makes a poor use of contextual information. The VDSR [10] network improves the efficiency of SRCNN by increasing the network depth from 3 to 20 convolutional layers. In order to increase the convergence speed, VDSR targets residuals instead of the ground truth image. However, since the residual is simply defined as the difference between the HR image and the upscaled LR image, it usually contains some sort of artifacts which misleads the training and degrades the performance of the HR image estimation. A deeply-recursive convolutional network (DRCN) is proposed in [11]. In order to increase the performance without introducing new parameters for additional convolutions, DRCN uses a very deep recursive layer (up to 16 recursions). The FSRCNN [9] improves the performance of the SRCNN, by adapting an hourglass shaped network. The shape of the network decreases the computational complexity by decreasing the number of convolutions. At the same time, since the number of layers is higher than SRCNN, a large amount of contextual information is used and the HR image reconstruction quality is increased. A residual dense network (RDN) is proposed in [29] to make full use of hierarchical features from the LR image. RDN extracts the abundant local features and uses a global feature fusion to learn both local and global features adaptively. In [30] the challenges of training an end-to-end deep convolutional neural network (CNN) for image SR is addressed via jointly training and ensemble of deep and shallow networks. A deep color-guided CNN framework is proposed in [31] for depth image SR. First, a data-driven filter for depth image is approximated. Then, a coarse-to-fine CNN is introduced to learn different kernel sizes. Finally, the color difference and spatial distance are fused for depth image SR. In [32] a combination of convolutional sparse coding (CSC) and deep CNN is proposed for image SR. First, CSC is used to extract image components and then, the deep CNN is used with a strong preference on residual components.

The existing deep learning-based methods show great performance for SR task. However, using LR image as input and HR image as output, decreases the training speed and performance. Residual learning tries to solve this problem by using the difference of upscaled LR and HR images as target. However, this type of training has two main issues. First, the differential image which is used as target in training, misleads the training and performance of the network due to including some noisy patterns. Second, since the size of input image is upscaled to HR image size, the computational complexity of network is increased. Proposed method solves these problems using the combination of skip connection and a deconvolutional layer. The skip connection, feeds the exact copy of the input to the deconvolutional layer and the remaining layers are forced to estimate residuals using a two-step training procedure. The main contributions of this work are as follows:

1) A new framework for image SR is proposed by combination of skip connection and deconvolutional layer. This framework decreases the complexity of network due to using the original LR image as input. In addition, the SR performance increases by letting deconvolutional layer to have access to the original LR image together with residual features.

2) A two-step training procedure is proposed to force the first part of the network to produce residual features. In first step, a conventional residual learning framework is used and the trained parameters are transferred into the main framework. These parameters are fine-tuned in the second step together with training the deconvolutional layer parameters.

3) Proposed framework is able to handle multiple scale factors by only fine-tuning the deconvolutional layer. This way, for new scale factors, the training is done very quickly compared to starting from the scratch.

## 3. Proposed Method

In this section, the design procedure of the proposed Deep Artifact-Free Residual Network (DAFR) is explained. Figure 1 shows the architecture of the proposed network. The input to our network is an LR image and the output is the upscaled HR image. Our model has two main parts: (1) feature extraction and enhancement and (2) image reconstruction. First part includes convolutional layers. In this part, the patches from the LR image are extracted. Then the representation of each LR patch is derived and mapped non-linearly into another representation which conceptually is the representation of target HR patch. The second part is a deconvolution layer which aggregates the enhanced features and up-samples them to reconstruct the final HR image. In the following, we explain the design details of the proposed network.

We use $(n + 2)$ layers in the feature extraction and enhancement part. The first layer includes 64 filters of support $c \times 5 \times 5$ where $c$ is the number of channels in the input image. All of the middle $n$ layers, include $m$ filters where the first one has size $64 \times 3 \times 3$ and the remaining have the size $m \times 3 \times 3$. The last convolutional layer has 64 filters of support $m \times 5 \times 5$.

The number of middle layers $n$ and the number of filters in each layer $m$, are two sensitive parameters which affect the performance and the computational complexity of our network. The effect of these parameters will be analyzed in section 4. By increasing $n$, the size of receptive filed will increase which allows us to exploit more contextual information from input image. A larger receptive field means that we are using more neighbor pixels to predict the image details. On the other hand, increasing the number of layers increases the computational complexity and makes it harder to train the network.

In order to make the size of output same as input after each convolution, we use zero-padding before convolutions which is shown to work for SR task [10]. After each convolutional layer, we use the Parametric Rectified Linear Unit (PReLU) [33] instead of Rectified Linear Unit (ReLU). The PReLU includes also the negative part of the coefficient and helps to avoid "dead features" [34] caused by zero gradients in ReLU. For the input signal $x$, the PReLU is defined as:

$$f(x) = \max(x, 0) + a \min(0, x) \qquad (1)$$

Where $f$ denotes the activation function and $a$ is the coefficient of the negative part which is learnable.

### 3.1. Artifact-Free Residual Learning

An HR image can be decomposed into low frequency and high frequency information. The low frequency information is nearly the same for LR and HR images. The high frequency information (image details) can be modeled as the difference between the HR image and the upscaled version of the LR image (residual image). In VDSR [10], the residual image is used as the target which should be predicted by the network. This method of learning allows to speed up training using high learning rates and deeper network structure. However, the drawback of this method is that the target residual image often includes some artifacts. In VDSR, a ground truth image ($Y$) is down-sampled using the scale factor $S$ to generate the LR image ($X$). Then, the residual image is the difference between the upscaled (using the Bicubic interpolation with same scale factor) version of the LR image ($\hat{X}$) and the ground truth image. If we denote the residual image with $R$, we have $R = Y - \hat{X}$. In the process of the generating upscaled LR images, some high frequency information is destroyed in the down-sampling step and Bicubic upscaling is not able to recover them. Therefore, the upscaled LR images lack some high frequency information of the corresponding ground truth images. The residual image includes all the lost high frequency information in the process of generating $\hat{X}$ from $Y$.

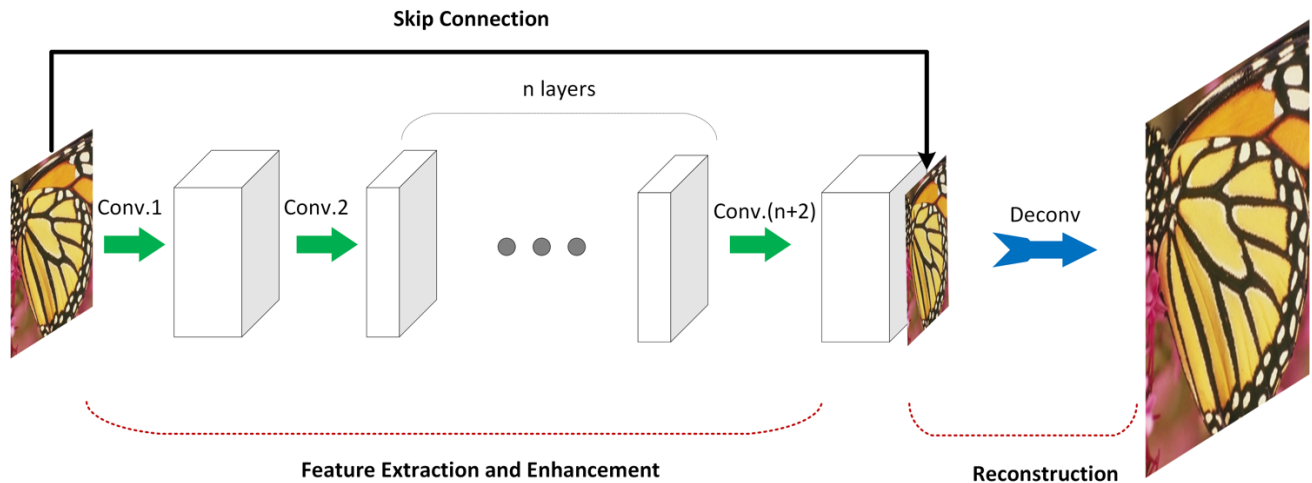

Figure 1. The Architecture of the Proposed Network. The input image is an LR image and the output is upscaled HR image. Feature extraction and enhancement part estimates the high frequency (residual) information by extracting the LR features and enhancing them. Skip connection feeds an exact copy of the LR image (low frequency information) to the deconvolutional layer. Deconvolutional layer uses both the low- and high-resolution information to reconstruct the HR image.

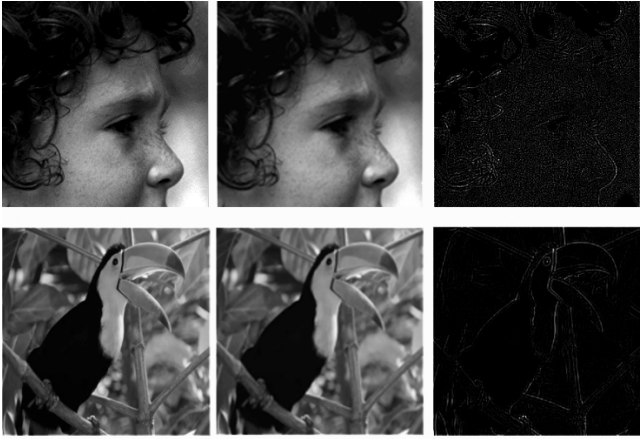

Figure 2. Noisy Residual Images Used for Training in VDSR

In figure 2, some ground truth, upscaled LR and corresponding residual images from set5 are shown in the first, second and third columns, respectively. As one can see, there are some strong features and some noisy patterns (artifacts) in the residual image. For example, in the *'face'* image (first row of Figure 2), the residual image contains some edge information and some noisy patterns. The edge information seems meaningful for the network. However, the noisy patterns (which may be related to the spot on the face or the image capturing noise) could be meaningless to the network, because there may be no relation between that patterns and the neighboring pixels. Therefore, these artifacts can mislead the network in training part and produce some noisy patterns in the reconstructed images.

Here, we propose a learning structure which trains the residual features and targets the ground truth image rather than noisy residual image. In the feature extraction and enhancement part of our network, the required features for the reconstruction of the residual image are extracted. In addition to the residual image (high frequency information), the low frequency information (within LR image) is necessary for image reconstruction. Since carrying the LR image is very inefficient for network training [10], we use skip (shortcut) connection to feed the input image directly to the reconstruction. This skip-connection has two main advantages in our network. First, the exact copy of the input signal can be used during target prediction. Second, with the availability of input image, the network capacity in feature extraction and enhancement part is forced to estimate the high frequency information. The skip connections [35, 36] are those skipping one or more layers and are realized by feedforward neural networks. In our proposed network, the skip connection simply performs identity mapping, and its output is concatenated to the outputs of the feature extraction and enhancement layers. Identity skip connection adds neither extra parameter nor computational complexity. The whole network can still be trained end-to-end by SGD with backpropagation, and can be easily implemented using common libraries without modifying the solvers [12].

In order to force the first part of the network to estimate the high frequency information, we separate the training into two steps. In the first step, we use a similar training procedure as [10] to train a network with $(n + 3)$ convolutional layers which reconstructs the residual image. The structure of this network is shown in Figure 3. For this training, we use high learning rates with adjustable gradient clipping [10, 37]. When the first step of training is saturated, in second step of training, we use the weights and biases of first $(n + 2)$ layers of this network as the initialization for first $(n + 2)$ layers of our proposed network (Figure 1). During the training, these parameters will be fine-tuned to produce more clean features. Inspired by [9], we use an *hourglass* shaped structure for feature extraction and enhancement, which is thick at the ends and thin in the middle. This shape is shown to reduce the complexity of the mapping features from the LR domain to the HR domain without loss of accuracy.

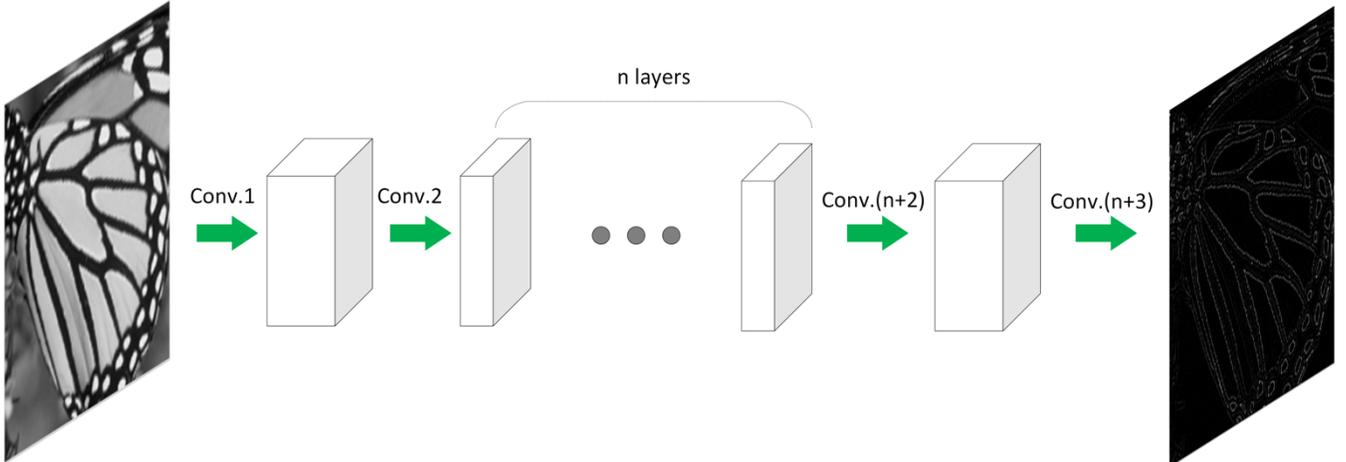

Figure 3. The structure of the residual network. Different from the proposed network, the input image is upscaled version of the LR image. The convolutional layers estimate the high frequency information (residual) of the input image. After training, the first (n+2) layers are transferred to the main network (Figure 1) as initial value.

### 3.2. Deconvolution Layer

The last layer of the proposed network is a deconvolution (transpose convolution) layer which aggregates the extracted features and input image (fed by the skip connection) to reconstruct the HR image with a set of deconvolution filters. From the viewpoint of input-output size, deconvolution is the inverse operation of the convolution. In the convolution, each filter is convolved with the input image (or feature map) with a stride $m$ and produces an output with $1/m$ size of the input. Therefore, in an inverse manner, for the deconvolution, the output will be $m$ times of the input. By setting the stride equal to the upscaling factor ($m = S$), the output of the deconvolution layer will be the reconstructed HR image.

### 3.3. Multi-scale SR

Most of the previous learning-based algorithms are trained for a single scale factor and work only with the specified scale. Therefore, for any other scale factors, a new model is required to be trained. The DAFR has an advantage of fast training across different scale factors over the previous algorithms. As we will discuss in the section 4, using the deconvolutional layer, we are able to quickly adapt the network for new scales by a small number of backpropagations.

## 4. Experimental Results

In this section, the performance of proposed method is evaluated. First, we explain the dataset used for training and testing. Then the training strategy is described. Next, the effect of sensitive parameters on the performance of network is investigated. Finally, we compare our method with some state-of-the-art SR methods.

### 4.1. Training

**Dataset** The 91 images from Yang et al. [6] is widely used in learning-based SR methods as training set. However, studies show that deep models benefit largely from big data and for the SR task, the 91 images are not enough to achieve the best performance. Therefore, similar to [8, 38], in addition to 91 images, we use a large set from the ILSVCR 2013 ImageNet dataset as our training dataset. Following SRCNN, FSRCNN and VDSR, three datasets 'Set5' [19], 'Set14' [39] and 'BSD200' [40] are used as test dataset.

**Training Strategy** For our main training structure (Figure 1), the original training images are downsampled by the scale factor $S$ to provide the LR images. Then, the LR images are cropped to provide a set of sub-images with size $f_{sub} \times f_{sub}$ pixels. The ground truth images are also cropped to prepare the corresponding HR sub-images with size $(Sf_{sub}) \times (Sf_{sub})$. The primary data of network training are these LR/HR sub-image pairs. For the residual structure (Figure 3), the original training images are downsampled by the scale factor $S$ and then interpolated by the Bicubic interpolation with same scale factor to provide interpolated LR image. Here, we use the $f_{sub-R} \times f_{sub-R}$-pixel sub-images cropped from interpolated LR images and corresponding ground truth images as training images.

Let $X$ denote a LR image (interpolated LR image for residual structure) and $Y$ denote the corresponding HR image. We aim to learn a mapping function $f$ to predict the value $\hat{Y} = f(X)$, where $\hat{Y}$ is the estimated HR image. Learning the mapping function $f$ requires to estimate network parameters $\Theta$ through minimization of the loss between the reconstructed image $f(X)$ and the corresponding ground truth HR image $X$. In our framework, we define a loss function to measure the error during training as follows:

$$L(\Theta) = \frac{1}{N}\sum_{i=1}^{N} \rho(f(X_i) - X_i) \qquad (2)$$

where N is the number of training samples and $\rho$ is a robust penalty function. Here we use Charbonnier penalty function $\rho(x) = (x^2 + 0.001^2)^\alpha$ [41] instead of mean square error which is proven to be robust enough to handle outliers. Training is carried out using mini-batch gradient descent based on back propagation [42].

As mentioned before, we adapt a two-step training procedure. First, we train the residual structure (Figure 3) from scratch with the 91-image dataset. We use batches of size 64 and set the momentum and weight decay parameters to be 0.9 and 0.0001 respectively. The initial learning rate is set to 0.1 and decreased by the factor of 10 every 20 epochs. We initialize the filter weights of each layer using a Gaussian distribution with zero mean and standard deviation 0.001 (and 0 for biases). When the training is saturated, in the second step of training, the weights and biases of the first ($n + 2$) layers of the residual network are transferred to the proposed network structure as initial value. The deconvolution layer uses 64 filters each of which have the spatial size $9 \times 9$. We use the bilinear kernel to initialize the filter weights of deconvolution layer and set the initial value of the bias to be zero. In the second step of training, the learning rate of the convolutional layers is set to be $10^{-5}$ and that of the deconvolution layer is $10^{-4}$.

### 4.2. Investigating of Different Setting

Here, we investigate the effect of the number of middle layers and the size of their feature map on the performance and complexity of our proposed network. Suppose a layer with $n_1$ input channels which includes $n_2$ filters with spatial size $n_1 \times f_1 \times f_1$. The number of parameters for this layer is $n_1 \times (f_1)^2 \times n_2$. Then the computational complexity of proposed network is

$$O\{(6784 + 2176m + (n-1) \times 9m^2)S_{LR}\} \qquad (3)$$

Where $S_{LR}$ denotes the size of LR image. In (3), the term within parentheses denotes the number of network parameters. Considering (3), the number of middle layers ($n$) and the number of filters in each layer ($m$) have a considerable effect on the complexity. Table 1 shows the network performance with the peak signal to noise ratio (PSNR) of the output image for different values of $n$ and $m$. In each case the number of network parameters is shown as an indication of computational complexity. From Table 1, it is obvious that $m$ has a major effect on the increasing complexity and a minor effect on the performance improvement. Based on the provided results, for the rest of the experiments in this paper, we use $n = 20$ and $m = 8$.

Table 1. The comparison of PSNR and parameters of different setting on set5.

|  | $m = 8$ | $m = 12$ | $m = 16$ |
|---|---|---|---|
| $n = 8$ | 32.89 (28224) | 32.93 (41968) | 32.95 (57728) |
| $n = 12$ | 33.22 (30528) | 33.24 (47152) | 33.24 (66944) |
| $n = 16$ | 33.29 (32832) | 33.31 (52336) | 33.30 (76160) |
| $n = 20$ | 33.35 (35136) | 33.36 (57520) | 33.32 (85376) |

### 4.3. Fast and Scale-Adaptive SR

Using deconvolution layer has several advantages in our work. First, we can use the original LR image (rather than its upscaled version) as input to our network. By upscaling LR image to the desired size (as in [8], [10], [11]) the computational complexity increases quadratically with the spatial size of the HR image. We use the deconvolution layer as the last layer for computational complexity of feature extraction and enhancement to be proportional to the spatial size of the original LR image. Besides, the deconvolution filters clean the artifacts of the extracted features and reconstruct a high-quality image. Finally, during our experiments, we find that for different scale factors, the convolution filters of the feature extraction and enhancement part are almost the same. Therefore, all convolutional layers in this part have similar behavior (like a complex feature extractor) for different scale factors and the information of scale factor only exists in the deconvolution layer. We use this property to accelerate the training of different scale factors by transferring the convolution filters. Specifically, we train our network for a scale factor. Then, for another scale factor we use the same convolution layers and only fine-tune the deconvolution layer.

In order to investigate the accuracy of this fine-tuning we perform the following experiment. We use the well-trained network for $\times 2$ as a basis and then fine-tune the deconvolution filter for the $\times 3$. The parameters of the convolution layers in $\times 2$ network are transferred to the $\times 3$ one. We also train another network for $\times 3$ using the method described in our training strategy. For simplicity, we call it training from scratch. Figure 4 shows the convergence curves for these two training methods. It is obvious that fine-tuning the deconvolutional layer for new scale factor results in a fast convergence and same performance as training from scratch.

### 4.4. Comparison with State-of-the-art

The qualitative and quantitative results of our algorithm in comparison to state-of-the-art is provided in this section.

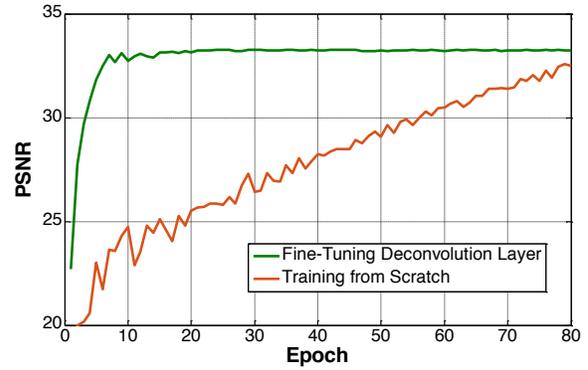

Figure 4. Convergence in Different Training Strategies for Upscale Factor 3 (Set5)

We compare our method with NARM [22], SRCNN [8], VDSR [10], DRCN [11] and FSRCNN [9]. NARM [19] is a nonlocal autoregressive algorithm which incorporates the image nonlocal self-similarity into sparse representation model. It is used in our comparison as a successful SC-based algorithm. SRCNN [8] is first deep-learning based algorithm which designed based on the general structure of the SC-based algorithms. As mentioned in section 2, VDSR, DRCN and FSRCNN are all extensions of SRCNN which try to improve the performance of SRCNN using deeper structures. The provided results for these algorithms are based on their released source code. Human visual system is more sensitive to image details than the color. Therefore, the majority of SR algorithms are applied on the luminance component only. Similar to conventional approaches, for color images SR, first the color image is transformed into the YCbCr space. The SR algorithms are applied only on the Y channel, while Cb and Cr channels are upscaled using Bicubic interpolation.

The quantitative results for the performance of our algorithm is shown in Table 2. In Table 2, the average PSNR values of the reconstructed HR images are tabulated for three scale factors (×2, ×3, ×4) and three test datasets. 'Set5' and 'Set14' datasets are usually used in evaluating the SR methods for benchmark. Besides, 'BDS200' dataset contains some challenging images. Results show that the proposed DAFR, outperforms previous methods in these datasets. Specifically, for scale factor ×3, the average PSNR gain achieved by DAFR are 0.08, 0.13 and 0.07 dB higher than the next best approach, DRCN [11], on three datasets. Note that since the DRCN is a very deep recursive network, the training and inference time of the proposed DAFR is considerably lower than it. The number of network parameters (NNP) is also included in Table 2. Note that the FSRCNN has the minimum NNP, however the proposed DAFR network outperforms the FSRCNN by a considerably large margin (*e.g.*, 0.89 dB on the Set5). Note that DRCN has the best performance after proposed method by NNP of 46080. It is worth noting that only the proposed method (DAFR) and FSRCNN use LR image as input. Therefore, the computational complexity of these methods is $O\{NNP \times S_{LR}\}$. However, SRCNN, VDSR and DRCN use upscaled LR image as input. Therefore, for scale factor $S$, the computational complexity of these methods is $O\{NNP \times S^2 \times S_{LR}\}$.

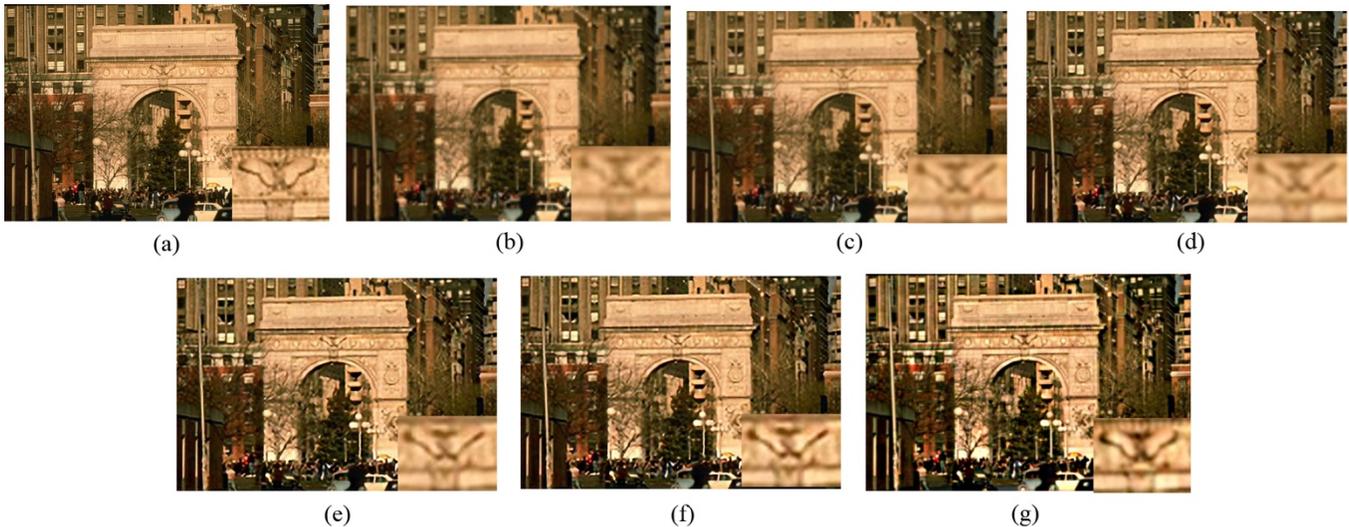

Figure 5. The Comparative Qualitative Results for scale factor ×3, (a) ground-truth image, reconstructed image by (b) bicubic, (c) NARM, (d) FSRCNN, (e) VDSR, (f) DRCN, (g) DAFR (proposed), method.

**Table 2.** The Quantitative Results for the Comparison of the Proposed Method with State-of-the-art.

| Algorithm | NNP | Scale Factor | Set5 | Set14 | BDS200 |
|---|---|---|---|---|---|
| Bicubic | - | ×2 | 33.66 | 30.23 | 29.70 |
|  |  | ×3 | 30.39 | 27.54 | 27.26 |
|  |  | ×4 | 28.42 | 26.00 | 25.97 |
| NARM [22] | - | ×2 | 36.01 | 31.89 | 30.81 |
|  |  | ×3 | 32.23 | 28.61 | 27.84 |
|  |  | ×4 | 29.98 | 27.01 | 26.32 |
| SRCNN [8] | 57184 | ×2 | 36.66 | 32.42 | 31.53 |
|  |  | ×3 | 32.75 | 29.28 | 28.47 |
|  |  | ×4 | 30.48 | 27.49 | 26.88 |
| DRCN [11] | 46080 | ×2 | 37.63 | 33.04 | 32.02 |
|  |  | ×3 | 33.82 | 29.76 | **28.99** |
|  |  | ×4 | 31.53 | 28.02 | 27.51 |
| VDSR [10] | 11520 | ×2 | 37.53 | 33.03 | 31.96 |
|  |  | ×3 | 33.66 | 29.77 | 28.91 |
|  |  | ×4 | 31.35 | 28.01 | 27.49 |
| FSRCNN [9] | 12464 | ×2 | 36.94 | 32.54 | 31.73 |
|  |  | ×3 | 33.06 | 29.37 | 28.55 |
|  |  | ×4 | 30.55 | 27.50 | 26.92 |
| DAFR (ours) | 35136 | ×2 | **37.68** | **33.12** | **32.10** |
|  |  | ×3 | **33.85** | **29.89** | 28.97 |
|  |  | ×4 | **31.72** | **28.09** | **27.62** |

Furthermore, the qualitative comparison results are shown in Figure 5. In order to produce these results, we have downsampled the ground-truth image by scale factor ×3 to produce the LR image, and then reconstructed the corresponding HR image by each algorithm. For each figure we have magnified a region of the reconstructed image to evaluate the quality of the reconstructed image. As one can see, the edges of the butterfly are distorted in other methods, however, the reconstructed image by proposed method has clean and vivid edges (specially the curves of wings). Note that based on the results of Table 2, DRCN has very close performance to our algorithm in terms of PSNR. However, using Charbonnier penalty as loss function enables us to produce more pleasant images.

## 5. Conclusion

In this work, we have proposed a deep convolutional network for image super-resolution. Using residual learning we are able to train a very deep network and use more contextual information for SR task. In traditional residual learning the target image is simply the difference between the HR and upscaled LR image which usually includes some artifacts. This misleads the network training and degrades its performance. We feed the LR image to image reconstruction step and this way we train our network with ground truth image where true features exist. Therefore, we use the benefits of residual learning without misleading the network in training procedure. We have used a two-step learning procedure to force our deep network for high frequency information extraction rather than carrying low frequency information. In experiments, we have shown that proposed method outperforms existing methods both quantitatively and qualitatively.

## References

[1] C. E. Duchon, "Lanczos filtering in one and two dimensions," *Journal of applied meteorology,* vol. 18, pp. 1016-1022, 1979.

[2] J. Sun, Z. Xu, and H.-Y. Shum, "Image super-resolution using gradient profile prior," in *Computer Vision and Pattern Recognition, 2008. CVPR 2008. IEEE Conference on*, 2008, pp. 1-8.

[3] K. I. Kim and Y. Kwon, "Single-image super-resolution using sparse regression and natural image prior," *IEEE transactions on pattern analysis & machine intelligence,* pp. 1127-1133, 2010.

[4] R. Timofte, V. De Smet, and L. Van Gool, "A+: Adjusted anchored neighborhood regression for fast super-resolution," in *Asian Conference on Computer Vision*, 2014, pp. 111-126.

[5] C.-Y. Yang and M.-H. Yang, "Fast direct super-resolution by simple functions," in *Proceedings of the IEEE international conference on computer vision*, 2013, pp. 561-568.

[6] J. Yang, J. Wright, T. S. Huang, and Y. Ma, "Image super-resolution via sparse representation," *IEEE transactions on image processing,* vol. 19, pp. 2861-2873, 2010.

[7] S. Schulter, C. Leistner, and H. Bischof, "Fast and accurate image upscaling with super-resolution forests," in *Proceedings of the IEEE Conference on Computer Vision and Pattern Recognition*, 2015, pp. 3791-3799.

[8] C. Dong, C. C. Loy, K. He, and X. Tang, "Image super-resolution using deep convolutional networks," *IEEE transactions on pattern analysis and machine intelligence,* vol. 38, pp. 295-307, 2016.

[9] C. Dong, C. C. Loy, and X. Tang, "Accelerating the super-resolution convolutional neural network," in *European Conference on Computer Vision*, 2016, pp. 391-407.

[10] J. Kim, J. Kwon Lee, and K. Mu Lee, "Accurate image super-resolution using very deep convolutional networks," in *Proceedings of the IEEE conference on computer vision and pattern recognition*, 2016, pp. 1646-1654.

[11] J. Kim, J. Kwon Lee, and K. Mu Lee, "Deeply-recursive convolutional network for image super-resolution," in *Proceedings of the IEEE conference on computer vision and pattern recognition*, 2016, pp. 1637-1645.

[12] A. Shah, E. Kadam, H. Shah, S. Shinde, and S. Shingade, "Deep Residual Networks with Exponential Linear Unit," in *Proceedings of the Third International Symposium on Computer Vision and the Internet*, 2016, pp. 59-65.

[13] K. He, X. Zhang, S. Ren, and J. Sun, "Deep residual learning for image recognition," in *Proceedings of the IEEE conference on computer vision and pattern recognition*, 2016, pp. 770-778.

[14] P. Sidike, E. Krieger, M. Z. Alom, V. K. Asari, and T. Taha, "A fast single-image super-resolution via directional edge-guided regularized extreme learning regression," *Signal, image and video processing,* vol. 11, pp. 961-968, 2017.

[15] C.-Y. Yang, C. Ma, and M.-H. Yang, "Single-image super-resolution: A benchmark," in *European Conference on Computer Vision*, 2014, pp. 372-386.

[16] D. Glasner, S. Bagon, and M. Irani, "Super-resolution from a single image," in *Computer Vision, 2009 IEEE 12th International Conference on*, 2009, pp. 349-356.

[17] J.-B. Huang, A. Singh, and N. Ahuja, "Single image super-resolution from transformed self-exemplars," in *Proceedings of the IEEE Conference on Computer Vision and Pattern Recognition*, 2015, pp. 5197-5206.

[18] A. M. Chaudhry, M. M. Riaz, and A. Ghafoor, "Super-resolution based on self-example learning and guided filtering," *Signal, Image and Video Processing,* vol. 13, pp. 237-244, 2019.

[19] M. Bevilacqua, A. Roumy, C. Guillemot, and M. L. Alberi-Morel, "Low-complexity single-image super-resolution based on nonnegative neighbor embedding," 2012.

[20] W. T. Freeman, T. R. Jones, and E. C. Pasztor, "Example-based super-resolution," *IEEE Computer graphics and Applications,* vol. 22, pp. 56-65, 2002.

[21] Y. Jianchao, J. Wright, T. Huang, and Y. Ma, "Image super-resolution as sparse representation of raw image patches," in *Proc. IEEE Conf. on Computer Vision and Pattern Recognition*, 2008, pp. 1-8.

[22] W. Dong, L. Zhang, R. Lukac, and G. Shi, "Sparse representation based image interpolation with nonlocal autoregressive modeling," *IEEE Transactions on Image Processing,* vol. 22, pp. 1382-1394, 2013.

[23] K. He, X. Zhang, S. Ren, and J. Sun, "Spatial pyramid pooling in deep convolutional networks for visual recognition," *IEEE transactions on pattern analysis and machine intelligence,* vol. 37, pp. 1904-1916, 2015.

[24] H. Nejati, H. A. Ghazijahani, M. Abdollahzadeh, T. Malekzadeh, N.-M. Cheung, K. H. Lee*, et al.*, "Fine-grained wound tissue analysis using deep neural network," *arXiv preprint arXiv:1802.10426,* 2018.

[25] S. Song, N.-M. Cheung, V. Chandrasekhar, and B. Mandal, "Deep Adaptive Temporal Pooling for Activity Recognition," *arXiv preprint arXiv:1808.07272,* 2018.

[26] T. Malekzadeh, M. Abdollahzadeh, H. Nejati, and N.-M. Cheung, "Aircraft Fuselage Defect Detection using Deep Neural Networks," *arXiv preprint arXiv:1712.09213,* 2017.

[27] T. Hoang, T.-T. Do, D.-K. Le Tan, and N.-M. Cheung, "Selective deep convolutional features for image retrieval," in *Proceedings of the 2017 ACM on Multimedia Conference*, 2017, pp. 1600-1608.

[28] M. Abdollahzadeh, T. Malekzadeh, and H. Seyedarabi, "Multi-focus image fusion for visual sensor networks," in *2016 24th Iranian Conference on Electrical Engineering (ICEE)*, 2016, pp. 1673-1677.

[29] Y. Zhang, Y. Tian, Y. Kong, B. Zhong, and Y. Fu, "Residual dense network for image super-resolution," in *Proceedings of the IEEE Conference on Computer Vision and Pattern Recognition*, 2018, pp. 2472-2481.

[30] Y. Wang, L. Wang, H. Wang, and P. Li, "End-to-end image super-resolution via deep and shallow convolutional networks," *IEEE Access,* vol. 7, pp. 31959-31970, 2019.

[31] Y. Wen, B. Sheng, P. Li, W. Lin, and D. D. Feng, "Deep color guided coarse-to-fine convolutional network cascade for depth image super-resolution," *IEEE Transactions on Image Processing,* vol. 28, pp. 994-1006, 2019.

[32] C. Xie, Y. Liu, W. Zeng, and X. Lu, "An improved method for single image super-resolution based on deep learning," *Signal, Image and Video Processing,* pp. 1-9, 2018.

[33] K. He, X. Zhang, S. Ren, and J. Sun, "Delving deep into rectifiers: Surpassing human-level performance on imagenet classification," in *Proceedings of the IEEE international conference on computer vision*, 2015, pp. 1026-1034.

[34] M. D. Zeiler and R. Fergus, "Visualizing and understanding convolutional networks," in *European conference on computer vision*, 2014, pp. 818-833.

[35] C. M. Bishop, *Neural networks for pattern recognition*: Oxford university press, 1995.

[36] K. Chatfield, K. Simonyan, A. Vedaldi, and A. Zisserman, "Return of the devil in the details: Delving deep into convolutional nets," *arXiv preprint arXiv:1405.3531,* 2014.

[37] R. Pascanu, T. Mikolov, and Y. Bengio, "On the difficulty of training recurrent neural networks," in *International Conference on Machine Learning*, 2013, pp. 1310-1318.

[38] C. Dong, C. C. Loy, K. He, and X. Tang, "Learning a deep convolutional network for image super-resolution," in *European conference on computer vision*, 2014, pp. 184-199.

[39] R. Zeyde, M. Elad, and M. Protter, "On single image scale-up using sparse-representations," in *International conference on curves and surfaces*, 2010, pp. 711-730.

[40] D. Martin, C. Fowlkes, D. Tal, and J. Malik, "A database of human segmented natural images and its application to evaluating segmentation algorithms and measuring ecological statistics," in *Computer Vision, 2001. ICCV 2001. Proceedings. Eighth IEEE International Conference on*, 2001, pp. 416-423.

[41] X. Xiang, M. Zhai, R. Zhang, Y. Qiao, and A. El Saddik, "Deep Optical Flow Supervised Learning With Prior Assumptions," *IEEE Access,* vol. 6, pp. 43222-43232, 2018.

[42] Y. LeCun, L. Bottou, Y. Bengio, and P. Haffner, "Gradient-based learning applied to document recognition," *Proceedings of the IEEE,* vol. 86, pp. 2278-2324, 1998.